\documentclass[aps,prl,twocolumn]{revtex4}
\usepackage{graphicx}
\begin{document}

%\preprint{}

\title{Enhancing the Superconducting Transition Temperature in the Absence of Spin Correlations in Heavy Fermion Compound CeIrIn$_5$}

\author{Shinji~Kawasaki$^1$}%
\author{Guo-qing~Zheng$^2$}
\author{Hiroki~Kan$^1$}%
\author{Yoshio~Kitaoka$^1$}%
\author{Hiroaki~Shishido$^3$}
\author{Yoshichika~\=Onuki$^3$}

\affiliation{$^1$Department of Materials Engineering Science, Graduate School of Engineering Science, Osaka University, Toyonaka, Osaka 560-8531, Japan\\$^2$Department of Physics, Okayama University, Okayama 700-8530, Japan\\$^3$Department of Physics, Graduate School of Science, Osaka University, Toyonaka, Osaka 560-0043, Japan}%

%\thanks{}
%\affiliation{}

\date{\today}

\begin{abstract}
We report on a pressure($P$)-induced evolution of  superconductivity and spin correlations in CeIrIn$_5$ via the $^{115}$In nuclear-spin-lattice-relaxation rate  measurements. We find that applying pressure suppresses dramatically the  antiferromagnetic fluctuations that are strong  at ambient pressure. At $P$ = 2.1 GPa, $T_{\rm c}$ increases to  $T_{\rm c}$ = 0.8 K that is twice $T_{\rm c}$($P$ = 0 GPa), in the background of Fermi liquid state. This is in sharp contrast with the previous case in which negative, chemical pressure (replacing Ir with Rh)  enhances magnetic interaction and increases $T_{\rm c}$. Our results suggest that multiple mechanisms work to produce superconductivity in the same compound CeIrIn$_5$. 
\end{abstract}

%\keywords{}

\maketitle
The  cerium (Ce)-based heavy-fermion compounds CeMIn$_5$ (M = Co, Rh, and Ir) discovered a few years ago provides a unique opportunity to  investigate the interplay between antiferromagnetism and superconductivity \cite{Hegger,Petrovic1,Petrovic2}. Among CeMIn$_5$, CeIrIn$_5$ and CeCoIn$_5$ show superconductivity at $P$ = 0 below $T_{\rm c}$ = 0.4 K and 2.3 K, respectively \cite{Petrovic1,Petrovic2}. Antiferromagnet CeRhIn$_5$ becomes superconducting  at relatively lower critical pressure $P_c$ $\sim$ 1.6 GPa and  yet exhibits a higher  $T_{\rm c}$ $\sim$ 2 K \cite{Hegger}. Measurements of nuclear-quadrupole-resonance (NQR)  \cite{Zheng,Mito,Kohori,KohoriEPJ}, thermal transport and heat capacity \cite{Movshovich} on CeMIn$_5$ found that the superconductivity is unconventional, with line-nodes in the superconducting gap function. NQR \cite{Zheng,Kohori,Mito,Yu} and inelastic neutron diffraction \cite{Bao} measurements also found strong antiferromagnetic spin fluctuations in the normal state. In addition, in CeIrIn$_5$, the antiferromagnetic spin fluctuations are found to be anisotropic\cite{Zheng}, namely, a magnetic correlation length $\xi_{\rm plane}$ within the tetragonal plane grows up more dominantly than $\xi_c$ along the $c$-axis associated with their two-dimensional crystal structure. 
The nuclear-spin-lattice-relaxation rate($1/T_1$) was found to follow the relation of $1/T_1T\propto 1/(T+\theta)^{3/4}$ with a small value of $\theta = 8$ K \cite{Zheng}. The same analysis was applied to CeCoIn$_5$ with resulting $\theta = 0.6$ K \cite{Yu,Yashima}. Note that $\theta$ is a measure to what extent the system is close to an antiferromagnetic quantum critical point (QCP)\cite{Lacroix}.  It was suggested that the difference in the value of $\theta$ between CeIrIn$_5$ and CeCoIn$_5$ may lead to the large difference in the value of $T_{\rm c}$. 

Moreover, substituting Ir to Rh in CeIrIn$_5$ increases $T_{\rm c}$ up to 1 K in CeRhIrIn$_5$\cite{Pagliuso}. It was found that this substitution acts as negative chemical pressure that increases the antiferromagnetic correlations \cite{Zheng2}. In fact, in CeRh$_{0.5}$Ir$_{0.5}$In$_5$, the enhanced superconductivity coexists microscopically with antiferromagnetic order that sets in at $T_N$ = 3 K \cite{Zheng2}. These results have naturally led to an expectation that superconductivity in CeMIn$_5$ is induced by antiferromagnetic correlations. However, it was found that applying hydrostatic pressure also increases $T_{\rm c}$ in  CeIrIn$_5$ \cite{Borth,Muramatsu}. A $T_{\rm c}^{max}\sim$ 1 K was found at around $P\sim$ 3 GPa \cite{Muramatsu}. The possible role of magnetic correlations in the increase of $T_{\rm c}$ under pressure in CeIrIn$_5$ is still an open question.

In this Letter, we report on the pressure-induced evolution of superconducting characteristics and antiferromagnetic spin fluctuations in CeIrIn$_5$ through  $T_1$ measurements. We found that the superconductivity with enhanced $T_{\rm c}$ under pressure in CeIrIn$_5$ is realized in the absence of antiferromagnetic spin fluctuations. Our results suggest that there are two mechanisms for superconductivity in the same compound  CeIrIn$_5$. We argue that the existence of multiple superconducting phase may be common in heavy fermion compounds.

Single crystals of CeIrIn$_5$ were grown by the self-flux method and moderately crushed into grains in order to allow rf pulses to penetrate easily into samples. To avoid crystal distortions, however, the grain's diameters were kept larger than 100 $\mu$m. CeIrIn$_5$ consists of alternating layers of CeIn$_3$ and IrIn$_2$ and hence has two inequivalent $^{115}$In sites per unit cell. The $^{115}$In-NQR measurements were made at the In(1) site \cite{Zheng} which is located on the top and bottom faces of the tetragonal unit cell. Since the position of In(1) site is crystallographically closer to Ce nucleus than that of In(2) site, it is suited to investigate the relationship between superconductivity and magnetic correlations. $^{115}$In-NQR measurement was made by a conventional saturation-recovery method. The $^{115}$In-NQR $T_1$ was measured at the transition of 2$\nu_{Q}$ ($\pm 3/2\leftrightarrow \pm 5/2$) above $T$ = 1.4 K, but at 1$\nu_{Q}$ ($\pm 1/2\leftrightarrow \pm 3/2$) below $T$ = 1.4 K. The hydrostatic pressure was applied by utilizing BeCu piston-cylinder cell, filled with Daphne oil (7373) as a pressure-transmitting medium. The value of pressure at low temperature was determined from the pressure dependence of the $T_{\rm c}$ value of Sn metal measured by a conventional four terminal method. For our pressure cells,  the spatial distribution in values of pressure $\Delta P/P$ is estimated to be $\sim 3 \%$ from a broadening in the linewidth of NQR spectrum\cite{Ykawasaki}.

%fig1
\begin{figure}[h]
\centering
\includegraphics[width=7cm]{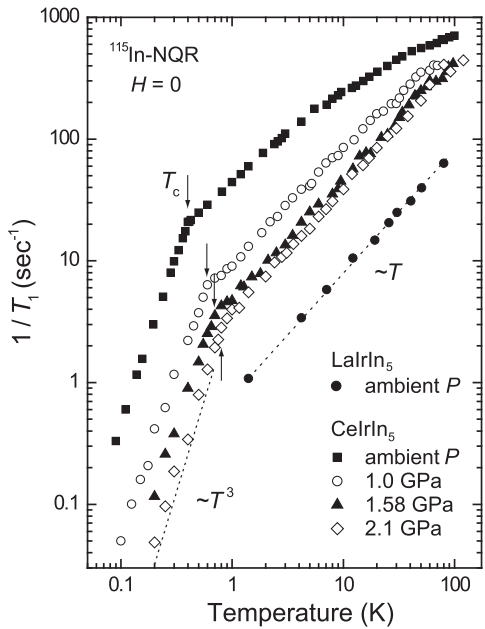}
\caption[]{The $T$ dependence of $^{115}(1/T_1)$ in CeIrIn$_5$ at $P$ = 0, 1.0, 1.58 and 2.1 GPa. The data for CeIrIn$_5$ at $P$ = 0 GPa and LaIrIn$_5$ are taken from Ref.\cite{Zheng}. Arrows indicate a superconducting transition temperature $T_{\rm c}$ at each pressure. The respective dotted lines indicate the behaviors of $1/T_1T$ = const. and $1/T_1\propto T^3$ at the normal and superconducting state.}
\end{figure}

Figure 1 shows the $T$ dependence of $^{115}$In-NQR $1/T_1$ for CeIrIn$_5$ measured at $P$ = 0, 1.0, 1.58 and 2.1 GPa. The data at $P$ = 0 GPa and for LaIrIn$_5$ are taken from Ref.\cite{Zheng}. Above $T_{\rm c}$, the Ce 4$f$ magnetic contribution to the relaxation rate for CeIrIn$_5$ is clear when comparing its value to $1/T_1$ measured in the non-magnetic LaIrIn$_5$. As reported in the previous work, the sudden decrease in $1/T_1$ at $P$ = 0 GPa at $T$ = 0.4 K indicated the onset of bulk superconductivity. Unconventional superconductivity was evidenced from the characteristic $T$ dependence of $1/T_1$ that exhibits no coherence peak just below $T_{\rm c}$ and follows the $T^3$ behavior well below $T_{\rm c}$ \cite{Zheng}. As pressure increases, $T_{\rm c}$ increases linearly and reaches $T_{\rm c}$ = 0.8 K which is twice the $T_{\rm c}$ at $P$ = 0 GPa. Note that the unconventional nature of superconductivity under pressure is evident from the $T$ dependence of $1/T_1$ below $T_{\rm c}$ as shown in Fig.2. 

In order to examine the pressure-induced evolution of superconducting characteristics in CeIrIn$_5$, $[T_1(T)^{-1}/T_1(T_{\rm c})^{-1}]$ versus $T/T_{\rm c}(P)$ is plotted in Fig.2. The line-node superconducting energy-gap model with $\Delta$ = $\Delta_0\cos \theta$ was applied to analyze the $1/T_1$ data below $T_{\rm c}$ with $\Delta_0/k_{\rm B}T_{\rm c}$ as a parameter. 

\[
\frac{T_1(T_{\rm c})}{T_{1}}=\frac{2}{k_{\rm B}T} \int \left( \frac{N_{\rm S}(E)}{N_0} \right)^2 f(E) [1-f(E)] dE,
\]
where $N_{\rm S}(E)/N_0$ = $E$/$\sqrt{E^2-\Delta^2}$ with $N_0$ being the density of state in the normal state and $f(E)$ is the Fermi distribution function. From fittings shown by solid line in Fig. 2, the pressure independent values of $\Delta_0/k_{\rm B}T_{\rm c}$ = 2.5 are obtained. Here, we assumed the residual density of state in the superconducting gap to be zero since clear $T^3$ behavior is observed down to 0.15 $T_{\rm c}$ in our sample.
This result shows that the coupling strength for the formation of Cooper pairs is almost the same in CeIrIn$_5$ regardless of the increase in $T_{\rm c}$. It is consistent with the results of specific heat measurements under pressure which showed that the relatively small specific heat jump at $T_{\rm c}$, $\Delta C/\gamma(T_{\rm c})\sim 0.8$, is almost independent of pressure\cite{Borth}. Here, $\gamma$ is the $T$-linear coefficient in electronic specific heat.
%fig2
\begin{figure}[h]
\centering
\includegraphics[width=7.5cm]{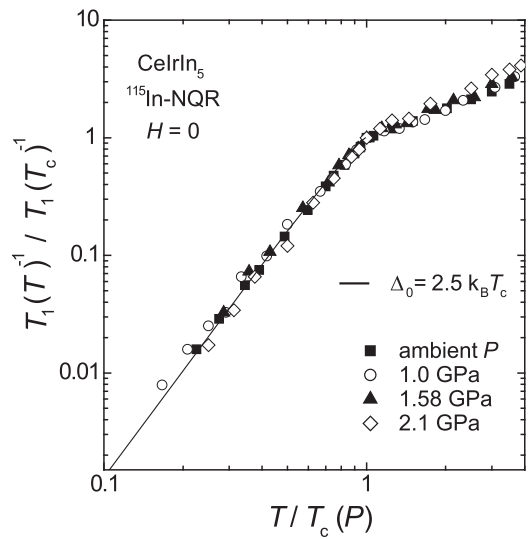}
\caption[]{Plot of $[T_1(T)^{-1}/T_1(T_{\rm c})^{-1}]$ versus $T/T_{\rm c}(P)$. Solid line indicates a calculation based on a unconventional superconducting model with a line-node gap assuming $\Delta_0$ = 2.5 $k_{\rm B}$$T_{\rm c}$ (see text). }
\end{figure}
What type of evolution in the electronic and magnetic properties under pressure increases $T_{\rm c}$ in CeIrIn$_5$? In order to gain an insight into this issue, we focus on the pressure-induced evolution of magnetic fluctuations in the normal state in CeIrIn$_5$. Figure 3 and its inset show the $1/T_1T$ versus $T$ plots in CeIrIn$_5$ at $P$ = 0, 1.0, 1.58 and 2.1 GPa and LaIrIn$_5$ at $P$ = 0 GPa in linear and logarithmic scales, respectively. At $P$ = 0 GPa, the $T$ dependence of $1/T_1T$ above $T_{\rm c}$ is well explained by the anisotropic antiferromagnetic spin fluctuations model \cite{Zheng,Lacroix}. As seen in Fig.3, the application of pressure markedly suppresses the antiferromagnetic spin fluctuations, bringing the system away from the antiferromagnetic QCP. As a result, a relation of $1/T_1T$ = const. becomes valid over $T = 1-100$ K at $P$ = 2.1 GPa without the development of antiferromagnetic spin fluctuations upon cooling. 
On the other hand, it should be noted that, as seen in Fig.4(b), the values of $1/T_1T$ and $\gamma\sim 0.25$ J/K$^2$mol\cite{Borth} for CeIrIn$_5$ at $P$ = 2.1 GPa are one order of magnitude larger than that for LaIrIn$_5$. These results indicate that strong electron correlation still plays a central role to enhance the values of $1/T_1T$ and $\gamma$, even though antiferromagnetic spin fluctuations disappear in CeIrIn$_5$ under pressure. This is in contrast to the case for CeCoIn$_5$ and CeRhIn$_5$ in which antiferromagnetic spin fluctuations enhance $1/T_1T$  upon cooling, even though each system is away from the antiferromagnetic QCP with applying pressure \cite{Yashima,KohoriEPJ}.  

%fig3
\begin{figure}[h]
\centering
\includegraphics[width=8cm]{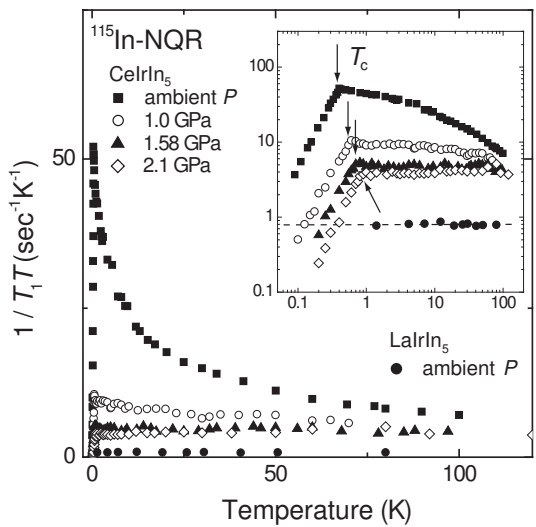}
\caption[]{The $T$ dependence of $1/T_1T$ at $P$ = 0, 1.0, 1.58 and 2.1 GPa. The data for CeIrIn$_5$ at $P$ = 0 GPa and LaIrIn$_5$ are taken from Ref.\cite{Zheng}. In the main figure and the inset, the data are plotted in linear and logarithmic scales, respectively.  Arrows and dotted lines indicate 
$T_{\rm c}(P)$ and a $T_1T$ = const. relation for LaIrIn$_5$, respectively.}
\end{figure}
The pressure dependencies of $\Delta_0/k_{\rm B}T_{\rm c}$ and $T_{\rm c}$ for CeIrIn$_5$ are summarized in Fig.4(a) together with those for CeCoIn$_5$ in the inset. In CeCoIn$_5$, the application of pressure also significantly suppresses $1/T_1$ that is dominated by antiferromagnetic spin fluctuations specific to the antiferromagnetic QCP. It shows very good agreement with specific heat measurements under pressure\cite{Sparn}. Although a jump in specific heat at $T_{\rm c}$ ($\Delta C/\gamma(T_{\rm c})\sim 5$) has a surprisingly large value at ambient pressure, indicative of a strong coupling superconductivity, this value shows a marked decrease against applying pressure \cite{Sparn}. Correspondingly, $1/T_1T$ is significantly suppressed as pressure increases \cite{Yashima}. Both results suggest that the application of pressure to CeCoIn$_5$ increases the heavy-fermion bandwidth due to the increase of hybridization between $f$ electrons and conduction electrons and eventually brings the system away from the antiferromagnetic QCP. As a result, the superconducting gap or $\Delta_0/k_{\rm B}T_{\rm c}$ in CeCoIn$_5$ decreases \cite{Sparn,Yashima} as shown in Fig. 4(a) inset. Noting that, in CeCoIn$_5$, the $1/T_1T$ is intimately enhanced upon cooling to $T_{\rm c}$, it is expected that antiferromagnetic spin fluctuations play a role in mediating the Cooper pairs even when the system is away from the antiferromagnetic QCP under pressure\cite{Yashima}. Nevertheless, the enhancement of $T_{\rm c}$ with applying pressure was suggested to be relevant to the increase in heavy-fermion bandwidth that is expected to make the lifetime of quasi-particles long enough. In this context, the Cooper pairs in CeCoIn$_5$ may originate from attractive interaction induced by antiferromagnetic spin fluctuations. It was hence argued that, in the presence of antiferromagnetic spin fluctuations, the value of $T_{\rm c}$ may be controlled by the combined effect of coupling strength for the Cooper-pair formation due to the closeness to the antiferromagnetic QCP and the heavy-fermion bandwidth in CeCoIn$_5$\cite{Yashima}. 

%fig4
\begin{figure}[h]
\centering
\includegraphics[width=7.5cm]{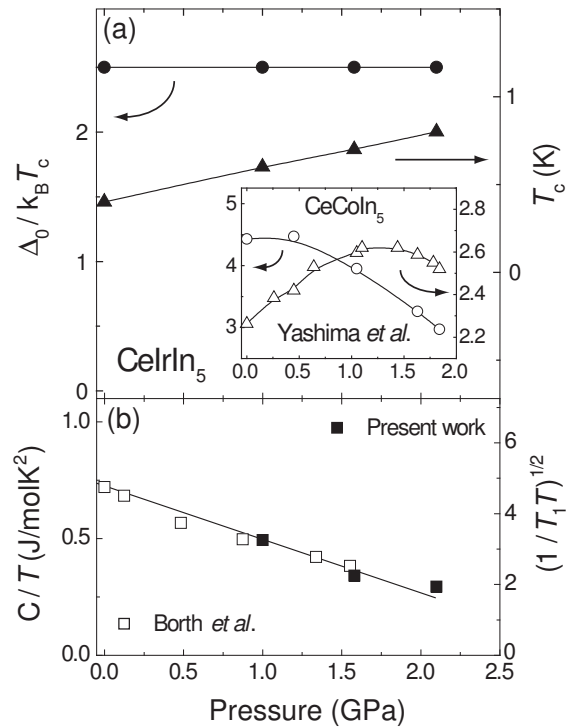}
\caption[]{ (a) The pressure dependence of $\Delta_0/k_{\rm B}T_{\rm c}$ and $T_{\rm c}$ in CeIrIn$_5$. The inset indicates those for CeCoIn$_5$ taken from the ref.\cite{Yashima}. Solid lines are eye guides.
(b) The pressure dependence of $\gamma$ \cite{Borth} and $(1/T_1T)^{1/2}$ in CeIrIn$_5$ just above $T_{\rm c}$($P$)(see text). Solid line is eye-guide.}
\end{figure}

This approach fails to account for the pressure dependence of $T_c$ in CeIrIn$_5$, however. Indeed, we have shown that the maximum of $T_c$ is realized without the development of antiferromagnetic spin fluctuations in the normal state as supported by the observation of $T_1T$ = const. law over two decades in the $T$ range above $T_{\rm c}$. The heavy-fermion bandwidth increases with pressure as corroborated by the fact that $\gamma$\cite{Borth} is scaled to $(1/T_1T)^{1/2}$ as seen in Fig.4(b). Here, the value of $(1/T_1T)^{1/2}$ is proportional to the density of state at the Fermi level. Therefore, the increase of $T_{\rm c}$ may be relevant to the increase of heavy-fermion bandwidth. 

In most Ce-based heavy-fermion compounds where the superconductivity appears either at $P$ = 0 GPa or in the neighborhood of antiferromagnetism, the antiferromagnetic spin fluctuations are expected to be responsible for the onset of unconventional spin-singlet superconductivity. In the antiferromagnetic spin fluctuations theory \cite{Moriya}, $T_{\rm c}$ takes a maximum value at the border of antiferromagnetism as observed in previous examples. In CeCu$_2$X$_2$ (X=Si and Ge), however, the maximum value of $T_{\rm c}$ appears far away from the antiferromagnetic QCP, which indicates that the low-lying antiferromagnetic spin fluctuations are not responsible for the formation of superconductivity \cite{Holmes}. In fact, it has been found that two superconducting phases exist in CeCu$_2$(Si$_{1-x}$Ge$_x$)$_2$ \cite{Yuan}. It has been suggested that one of superconductivity (SC-I) is induced by strong antiferromagnetic spin fluctuations on the verge of antiferromagnetism and the other (SC-II) by valence instability of localized Ce-4$f$ electrons since the system is far away from the antiferromagnetic QCP \cite{Onishi,Miyake,Holmes}. Markedly, the higher $T_{\rm c}$ takes place in the SC-II. 

Approaches based on spin-fluctuations theories also fail to account for the existence of spin-triplet superconductivity in Sr$_2$RuO$_4$ where two-dimensional Fermi-liquid state is realized with the strong electron correlation as confirmed by the $T_1T$ = const. law. It has been proposed that the on-site electron correlation induces various types of unconventional superconductivity through the momentum dependence of quasi-particle interaction, which originates from the many-body effect \cite{Yanase}. Interestingly, this scenario can predict $d$-wave and $p$-wave superconductivity near half-filling and away from half-filling respectively without involving spin-fluctuations as the key-mechanism for pairing\cite{Yanase}.

An important fact revealed by the present experiment is that superconductivity in CeIrIn$_5$ is robust over a wide pressure range where the antiferromagnetic spin fluctuations are absent. This may be due to either charge valence instability or, on-site coulomb interactions. Further experiments are required at this stage to clarify this issue. 
  
In summary, we have presented the unique characteristics of superconductivity and its relation to antiferromagnetic spin fluctuations in the heavy-fermion superconductor CeIrIn$_5$ through  $^{115}$In-NQR measurements under pressure.   The application of external pressure rapidly suppresses antiferromagnetic spin fluctuations that are strong at ambient pressure. At $P$ = 2.1 GPa where a $T_1T$ = const. law is valid over $T = 1-100$ K, $T_{\rm c}$ increases up to $T_{\rm c}$ = 0.8 K which is twice the $T_{\rm c}$ at $P$ = 0 GPa. Our results indicate that another superconducting phase exists  in the absence of antiferromagnetic spin fluctuations, in addition to the superconducting phase with $T_{\rm c}^{max}$ = 1 K that coexists with antiferromagnetism. The present system bears some similarity with another proto-type heavy fermion compound CeCu$_2$Si$_2$, and suggests that the existence of multiple superconducting phase may be common in heavy fermion compounds. 

We thank M.~Yashima and H.~Kotegawa for experimental assistance. This work was supported in part by Grants-in-Aid for Creative Scientific Researchi15GS0213), MEXT and The 21st Century COE Program of the Japan Society for  the Promotion of Science. S. K. has been supported by a Research Fellowship of the Japan Society for the Promotion of Science for Young Scientists. G.-q.Z acknowledges MEXT Grant No.16340104.

\end{document}